\documentclass[twocolumn, amsmath,amssymb, aps]{revtex4-2}

\usepackage{graphicx}
\usepackage{dcolumn}
\usepackage{bm}
\usepackage{array}
\usepackage{comment}
\usepackage{todonotes}
\usepackage{subcaption}
\usepackage[percent]{overpic}

\begin{document}

\title{Hydrodynamic Mode Coupling: Effects of density variations in nanoscale channel flows.}

\date{\today}
\author{L. Heitmeier}
\email{heitmeier@ruc.dk}
 \affiliation{Institute of Frontier Materials on Earth and in Space, German Aerospace Center, Cologne, Germany}
  \affiliation{Department of Physics, Heinrich-Heine Universität Düsseldorf, Universitätsstraße 1, 40225 Düsseldorf, Germany}
  \affiliation{``Glass and Time'', IMFUFA, Department of Science and Environment, Roskilde University, Postbox 260, DK-4000 Roskilde, Denmark}
\author{J. S. Hansen}
 \email{jschmidt@ruc.dk}
\affiliation{``Glass and Time'', IMFUFA, Department of Science and Environment, Roskilde University, Postbox 260, DK-4000 Roskilde, Denmark}

\begin{abstract}
    We apply a modal framework for investigating the effect of density variations on gravity-type driven flows at the nanoscale. Using eigenfunction decomposition of the density and acceleration fields, each shear-pressure mode is separated into a homogeneous contribution and an inhomogeneous contribution determined by the Fourier amplitudes of the density and applied acceleration. This decomposition provides a direct means of identifying how density variations and external forcing couple and govern the flow behavior. We first revisit the Poiseuille flow and show that for channel heights larger than the characteristic intermolecular distance the homogeneous contribution dominates the long wave length (small wave vector) response, consistent with previous simulation results. In contrast, for sinusoidally driven flow, selective excitation of acceleration modes can produce the opposite behavior, with the inhomogeneous contribution dominating the long wave length response. The results show that the effect of the density variations depends on the specific flow; specifically the detailed mode coupling between the acceleration and density fields. The framework presented here provides a direct systematic approach for understanding and predicting the flow depending on the applied acceleration. 
    
	\begin{description}
	\item[Keywords]
		Nanoconfinement, Hydrodynamics, Density-force coupling, Molecular dynamics  
	\end{description}
	
\end{abstract}

\maketitle
The continuum hypothesis assumes that quantities like density, velocity, and energy 
can be represented mathematically as field variables even if matter ultimately consists of  
discrete entities like atoms and molecules \cite{lautrup:book:2005}. This hypothesis together
with the mass and momentum balance equations, and a linear constitutive equation, namely, Newton's viscosity
law, give the famous Navier-Stokes (NS) equation that describes fluid flows. 
For systems on the nanoscale the characteristic system length scale  
becomes comparable with the size of the fluid molecules, and the validity of the 
NS equation is immediately questionable. However, several molecular dynamics studies  
have now established that the NS equation is indeed applicable on the nanoscale, see e.g. Refs. 
\cite{koplik:pfa:1989,travis:pre:1997,hansen:mn:2006,hansen:pre:2011:2,hansen:book:2022}. 

As pointed out by Bitsanis et al. four decades ago \cite{bitsanis:jcp:1987}, this is surprising 
considering that the fluid density features very large variations throughout a nanoscale 
channel; this is due to the fluid molecule layering. At first thought, this layering should affect
the local transport properties significantly, nevertheless, the NS prediction for the flow is
satisfactory using the constant macroscopic viscosity that enters Newton's viscosity law. 
To account for the absent effects from the density variations 
Bitsanis et al. proposed a local averaged density model (LADM) 
\cite{bitsanis:jcp:1987,bitsanis:jcp:1988}, where the transport properties at a point are
determined by a local averaged density rather than the density at the point. This averaged 
density features much smaller variations than the corresponding point-wise density suppressing the
density variation effect. While predicting the flow profile for Couette and Poiseuille flows 
this model does not account for the reduced stress observed in simulations for constant
density \cite{hansen:book:2022}. Cadush et al. \cite{cadusch:jphysa:2008} applied generalized 
hydrodynamics in order to account for the small effect of the density variations, however, here 
the viscosity kernel support is unclear, also at the wall-fluid interface, and as with the LADM  
ad-hoc choices must be made. Thus, it is still an open question why density variations have so small effect on the flow.

Knudsen et al. \cite{knudsen:pof:2025} and Heitmeier et al. \cite{heitmeier:pof:2025} recently 
reported an opposite effect of large length scale excitations of fluid velocity 
modes for a steady-state nanoscale flow driven 
by a single mode sinusoidal force. It was conjectured that the additional modes are induced by density 
and wall-fluid shear modes, however, it remains to show exactly how these large scale modes emerge and  
how the modes depend on the external driving force. 

We therefore revisit the continuum theory predictions of nanoscale fluid flows. We will do this by 
expanding the field variables in terms of eigenfunction modes. Through molecular dynamics
simulations this enables an investigation of how the different density modes couples to the modes in the 
external driving force. It is important to mention that Dalton et al. performed an in-depth spectral analysis of sinusoidal flows, where the density variation can be induced by a synthetic driving force \cite{dalton:pre:2013, dalton:pre:2015}; here we investigate confined flows using direct nonequilibrium simulations, which enables us to investigate experimental relevant situations like e.g. the Poiseuille flow. This also ensures that the density variation is a result of confining walls and not an applied synthetic force. 

\begin{figure}
	\centering
	\includegraphics[scale=0.45]{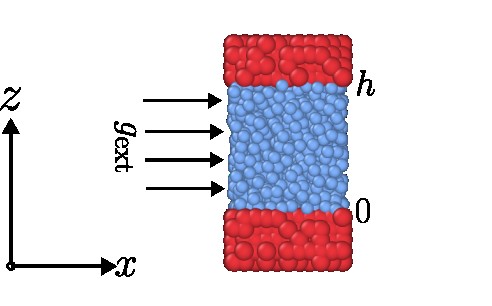}
	\caption{\label{fig:snapshot}
		Simulation snapshot of the nanoscale slit pore in case of the Lennard-Jones system. 
        Blue particles are liquid particles, and the red particles are the wall particles that are tethered to
		body-centered cubic lattice sites. Arrows indicates the external acceleration for the
		Poiseuille flow. The $y$-axis points normal to the page. 
	}
\end{figure}

\emph{Systems and Theory} We investigate fluids confined between two parallel 
walls, such that the $z$-direction is the direction of confinement, and the $xy$-plane 
is infinite in extend; Fig. \ref{fig:snapshot} illustrates the system. 
An external acceleration, $g_\text{ext}=g_\text{ext}(z)$, acts on the fluid in the $x$-direction giving 
a non-zero flow in this direction. For sufficiently small external accelerations, i.e., small Reynolds numbers, the momentum balance equation can be written as 
$\rho \partial_t u = \rho g_\text{ext} - \partial_z P_{xz}$. Here $\rho$ is the local density, 
$u$ is the fluid velocity $x$-component and $P_{xz}$ is the $xz$-component of the pressure
tensor which we denote the shear pressure. Notice that $\rho g_\text{ext}$ is a body force density. 
The viscous heating is conducted away through the walls and due to the low Reynolds number 
we assume a constant system temperature. In the steady-state the momentum balance equation reads 
\begin{align}
	\label{eq:stmombalance}
	\frac{\mathrm{d}P_{xz}}{\mathrm{d}z} = \rho g_\text{ext} \, .
\end{align}
Both $\rho$ and $g_\text{ext}$ are, in general, position dependent. 
Equation (\ref{eq:stmombalance}) is simply a balance between surface and body density 
forces, and we therefore consider this a fundamentally true relation for the steady-state. 
It does, however, not provide the fluid velocity, and to this end we apply 
Newton's constitutive model, $P_{xz} = -2\eta_0 \dot{\gamma}$, where $\eta_0$ is the shear viscosity
and $2\dot{\gamma} = \mathrm{d}u/\mathrm{d}z$ is the strain-rate.  The viscosity 
$\eta_0$ is a state dependent material property, i.e., if $\eta_0$ is constant we 
also assume, strictly, that the density is constant, $\rho=\rho_0$.
Application of Newtons's model to Eq. (\ref{eq:stmombalance}) gives the NS equation for
the steady-state  
\begin{align}
    \label{eq:2ndorderproblem}
	\frac{\mathrm{d}^2u}{\mathrm{d}z^2} = -\frac{1}{\eta_0} \rho_0 g_\text{ext}\, .
\end{align}
This local (and linear) model will fail if the strain rate varies on sufficiently 
small length scales \cite{travis:pre:1997,todd:prl:2008} - typically on the order of 
2-3 nm for many fluid systems \cite{hansen:book:2022}. 

Consider the case of no-slip boundaries, i.e., homogeneous Dirichlet boundaries $u(0)=u(h)=0$.
We seek solutions to Eq. (\ref{eq:2ndorderproblem}) with these boundaries  
in terms of an eigenfunction series; the eigenfunctions $\phi_n$ are found 
from the Sturm-Liouville problem $\mathcal{L}[\phi_n]=-k_n \phi_n$, where $k_n$ is the eigenvalue and where we apply the Dirichlet boundaries $\phi_n(0)=\phi_n(h)=0$ and $\mathcal{L} = \mathrm{d}^2/\mathrm{d}z^2$.  Including the zero solution we have $\phi_n = \sin(k_n z)$ with $k_n = n\pi/h$, $n \in \mathbb{N}$, that is, we have the solution $u = \sum_{n=1}^\infty u_n \sin(k_n z)$. This is just a Fourier sine series on $0 \leq z\leq h$ where $k_n$ is the wave vector. We will refer to the $n$'th term in the series as the $n$'the mode, and the coefficients as amplitudes. 

The left-hand side of Eq. (\ref{eq:2ndorderproblem}) is also a Fourier sine 
series since the second order derivative is
\begin{align}
    \label{eq:lhside}
    \frac{\mathrm{d}^2u}{\mathrm{d}z^2} = - \left(\frac{\pi}{h}\right)^2\sum_{n=1}^{\infty} 
	u_n n^2 \sin\left(k_n z\right) \, .
\end{align}
This motivates an expansion of the external
acceleration in terms of a sine Fourier series; for constant density we then have 
$\rho g_\text{ext} = \rho_0 \sum_{n=1}^\infty g_n \sin(k_n z)$. Comparing term-wise with Eq. (\ref{eq:lhside}) 
we obtain an expression for the velocity amplitudes in terms of $\rho_0$ and $g_n$ for $n>0$
\begin{align}
	\label{eq:un_homogen}
	u_n = \frac{\rho_0 g_n}{\eta_0 k_n^2} \ \ \text{(Hom. case)}.
\end{align}
From Newton's model, the shear pressure is seen to be a Fourier cosine series 
$P_{xz} = \sum_{n=0}^\infty P_{n} \cos(k_n z)$ with amplitudes 
\begin{align}
	\label{eq:shearHom}
	P_{n} = -\rho_0 g_n/k_n \ , 
\end{align}

From this knowledge we can proceed to the general case where the density varies. The density 
can be expanded as a cosine series $\rho = \sum_{n=0}^\infty \rho_n
\cos(k_n z)$. Then, the product on the right-hand side of Eq. (\ref{eq:stmombalance}) reads 
\begin{align}
	\rho g_\text{ext} &= \sum_{l=0}^\infty\sum_{m=1}^\infty
	\rho_lg_m \cos\left(k_lz\right)\sin\left(k_m z\right) \nonumber \\
	&= \frac{1}{2} \sum_{l=0}^\infty\sum_{m=1}^\infty\rho_lg_m
	\left(\,
		\sin\left(k_{m-l} z\right) + \sin\left(k_{m+l}z\right) \, \right) \label{eq:rhside}
\end{align}
which is a sine series as expected from above. Differentiation of the cosine series 
shear pressure we can again compare term-wise in Eq. (\ref{eq:stmombalance}) and obtain for $n>0$
\begin{align}
	\label{eq:shearPressGen}
	P_{n} = -\frac{1}{2 k_n} (2\rho_0 g_n + a_n)
\end{align}
where
\begin{align}
	\label{eq:an}
	a_n = \sum_{i=1}^{n-1}\rho_i g_{n-i}  + \sum_{j=1}^\infty (\rho_jg_{n+j} - \rho_{n+j}g_j) \, .
\end{align}
The first term on the right-hand side in Eq. (\ref{eq:shearPressGen}) recaptures the 
homogeneous case. The second term is a measure of the density-acceleration 
coupling for varying density at a given wave vector $k_n$, or equivalently at, 
a given length scale, $l_n = 2\pi/k_n$. Comparing the homogeneous part, $\rho_0 g_n$, 
and inhomogeneous part, $a_n$, of the mode $n$ we can now investigate the 
effect of coupling of density and acceleration modes. The shear pressure is the 
main quantity of interest rather than the fluid streaming velocity directly.  
It is also important to notice that from the pre-factor in Eq. (\ref{eq:shearPressGen}), 
we expect that both the homogeneous and inhomogeneous terms go to zero as
$k_n \rightarrow \infty$.

\emph{Molecular simulations} To investigate the mode coupling we perform a series 
of standard direct nonequilibrium molecular dynamics simulations of a Lennard-Jones liquid, see e.g. Ref. \cite{allen:book:1989,rapaport:book:1995}, and a model butane liquid \cite{ryckaert:fdcs:1978}. 
The liquids are confined between two atomistic walls, where the wall atoms are allowed to vibrate 
around the lattice sites of a wall crystal structure. The wall atoms are thermostated 
\cite{nose:molphys:1984,hoover:pra:1985,sadus:book:1999} such that the system temperature is constant. 
An acceleration is imposed on the fluid particles; the magnitudes of the acceleration are kept sufficiently low in order to achieve low Reynolds number and low viscous heating. 
In the following we will express quantities in reduced molecular dynamics simulation units, e.g., 
length will be in units of a Lennard-Jones particle diameter. Additional details 
can be found in the supplementary material. Figure \ref{fig:snapshot} shows a snapshot a Lennard-Jones simulation. 

The amplitudes for the density is found numerically using the Euler-Fourier equation \cite{boyce:book:1997} $\rho_n = 2/h \int_0^h \rho(z)\cos(k_n z) \mathrm{d} z$, where $\rho$ is calculated using a straight-forward bin method; 
details is given in the supplementary material. 
Figure \ref{fig:pois} (a) shows the no-flow equilibrium density profile for the Lennard-Jones system, 
where the channel height is $h = 7.8$ particle diameters. The characteristic density variations 
are clearly observed with the period of approximately one particle diameter as is well-known. The inset shows the corresponding amplitude spectrum, $\rho_n$, featuring 
many excited modes, and, as expected, the bulk density mode $\rho_0$ is dominant. Notice that for $k_n \approx 2\pi$  (vertical dashed line), corresponding to $l_n \approx 1$, the amplitude increases
abruptly. This is a finger print of the layering with period of around unity as stated above.   
The density profile is independent of the acceleration applied here and the profile in the flow case 
resembles the equilibrium profile.

\begin{figure}[h!]
    \hspace{0.02\textwidth}
    \begin{overpic}[scale=0.41]{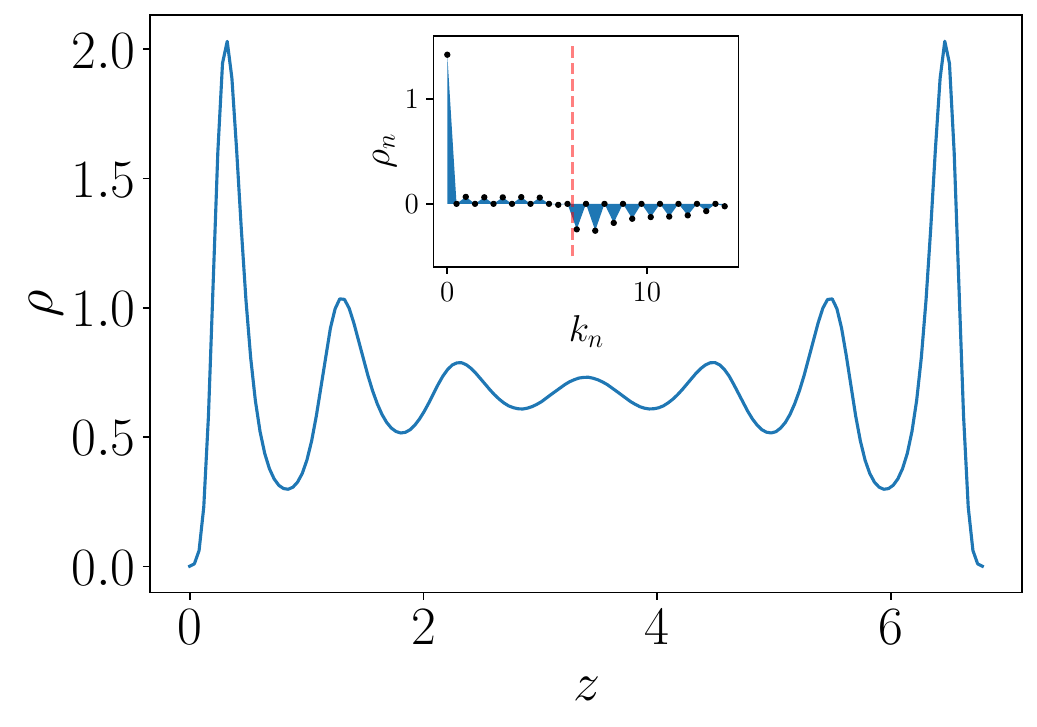}
        \put(-7,70){\makebox(0,0)[lt]{\large\textbf{a)}}}
    \end{overpic}
    \hspace{0.02\textwidth}
    \begin{overpic}[scale=0.5]{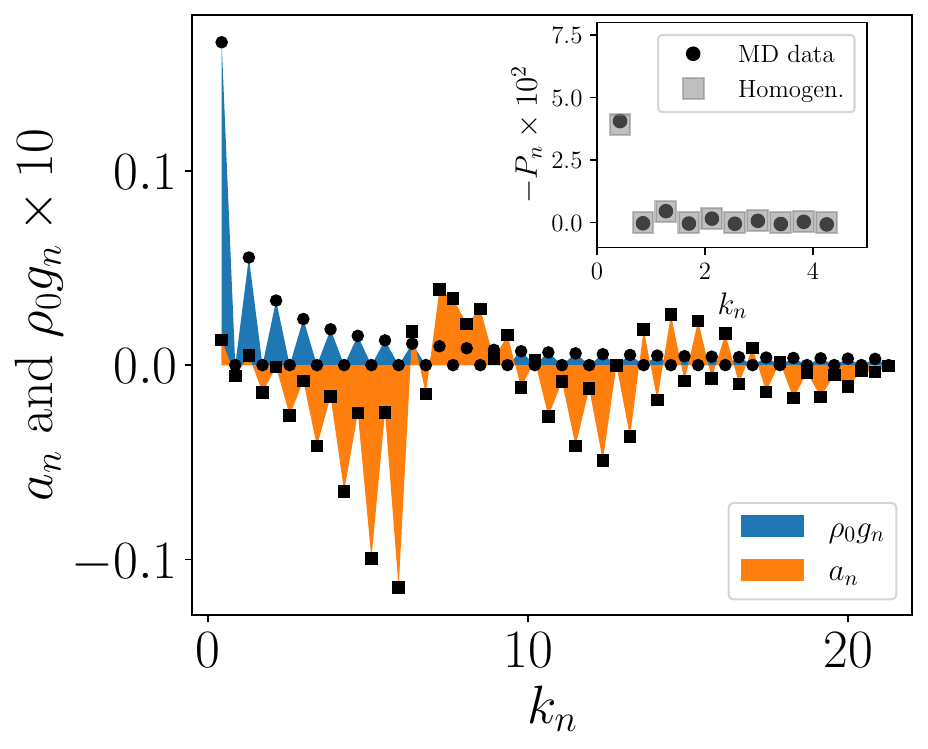}
        \put(-4,80){\makebox(0,0)[lt]{\large\textbf{b)}}}
    \end{overpic}
    \hspace{0.02\textwidth}
    \begin{overpic}[scale=0.5]{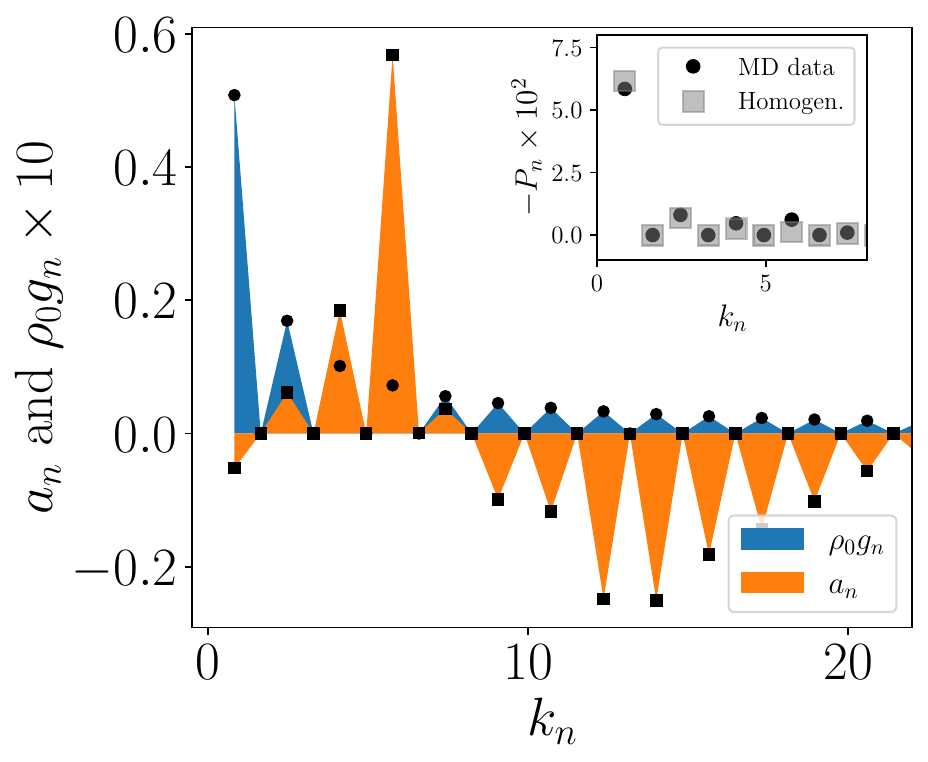}
        \put(-4,75){\makebox(0,0)[lt]{\large\textbf{c)}}}
    \end{overpic}

    \caption{\label{fig:pois}
    (a) Density profile for $h = 7.8$. Inset shows the corresponding amplitude spectrum;
    vertical line highlights the wave length $l_n \approx 1$.
    (b) $h=7.8$; homogeneous part, $\rho_0 g_n$, and inhomogeneous part, $a_n$, of the
    spectrum for the shear stress. Inset shows $P_n$ using Eqs. (\ref{eq:shearHom}) and (\ref{eq:shearPressGen}).
    (c) Same as (b), but where $h=3.8$.
    }
\end{figure}

\textit{Poiseuille flow} 
The Poiseuille flow is driven by $g_\text{ext} = G_0 1_{[0;h]}(z)$, where 
$1_{[0;h]}$ is the indicator function on the interval $[0;h]$. The amplitudes for
$g_\text{ext}$ are found analytically from the Euler-Fourier integral to $g_n = 4G_0/\pi n$, 
when $n$ is odd and zero otherwise. Thus, in the homogeneous case we get for $n=1,3, \ldots$,  
$\rho_0 g_n = 4\rho_0 G_0/\pi n = 4\rho_0G_0h/k_n$, and the velocity and shear pressure spectra are 
$u_n = 4h^2\rho_0G_0/\eta_0(n\pi)^3$ and $P_n = - 4\rho_0G_0h/(\pi n)^2$, respectively. 
Due to symmetry $P_0$ is zero.  

Figure \ref{fig:pois} (b) shows $\rho_0g_n$ and $a_n$ as functions of wave vector  
for the case where $h=7.8$. First, it can be seen that the homogeneous part 
$\rho_0g_n$ dominates at small wave vector. As $k_n$ increases 
the magnitude for the inhomogeneous part increases and shows a maximum at around 
$k_n \approx 6$ or length scale of approximately one particle diameter. This maximum agrees with the
abrupt local amplitude increase for density and we assign the particle layering to be the phenomenon behind the  
maximum in $a_n$. From Eq. (\ref{eq:shearPressGen}) $P_{n} \propto 1/k_n$, hence, for
channel heights that are sufficient large compared to the layering length scale
the small wave vector homogeneous modes will dominate the shear 
pressure spectrum and thus the fluid dynamics even though large wave vector 
mode coupling between the external acceleration and density exists. For the dimensionless case here this  
implies that $h$ must be sufficiently greater than one for the inhomogeneous part to be insignificant. 
The inset in Fig. \ref{fig:pois}(b) plots the amplitudes of the shear pressure using Eqs. (\ref{eq:shearHom}) and 
(\ref{eq:shearPressGen}); this confirms that the dynamics is indeed dominated 
by the homogeneous part $\rho_0 g_n$. 

The number of small wave vector homogeneous modes that can be excited before $k_n
\approx 6$ is reduced in small channel height $h$. Figure \ref{fig:pois} (c) shows the same as (b), 
but for $h=3.8$. Clearly, the homogeneous part dominates for the two smallest wave vectors as expected, however, 
for $n>2$ the inhomogeneous will dominate the $n$-mode shear pressure. The inset shows that this results in a significant difference in the shear pressure amplitude for $k_n \approx 2\pi$ between the homogeneous case and inhomogeneous case. 

\begin{figure}
     \hspace{0.02\textwidth}
    \begin{overpic}[scale=0.41]{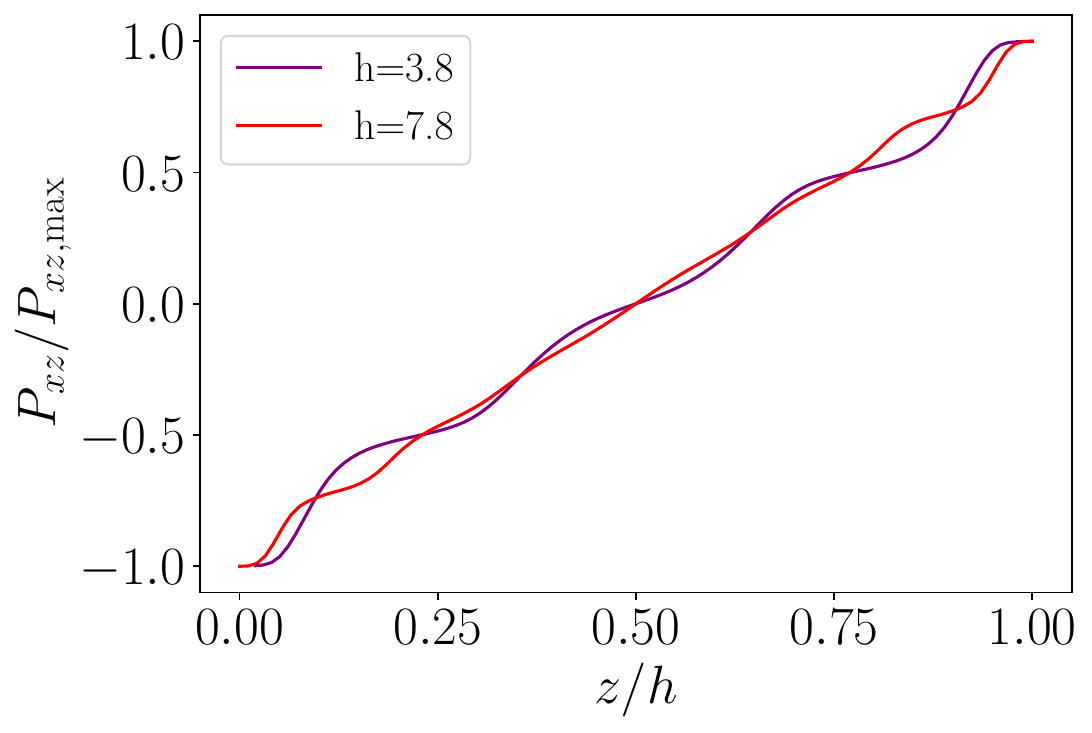}
        \put(-7,70){\makebox(0,0)[lt]{\large\textbf{a)}}}
    \end{overpic}

    \hspace{0.02\textwidth}
    \begin{overpic}[scale=0.41]{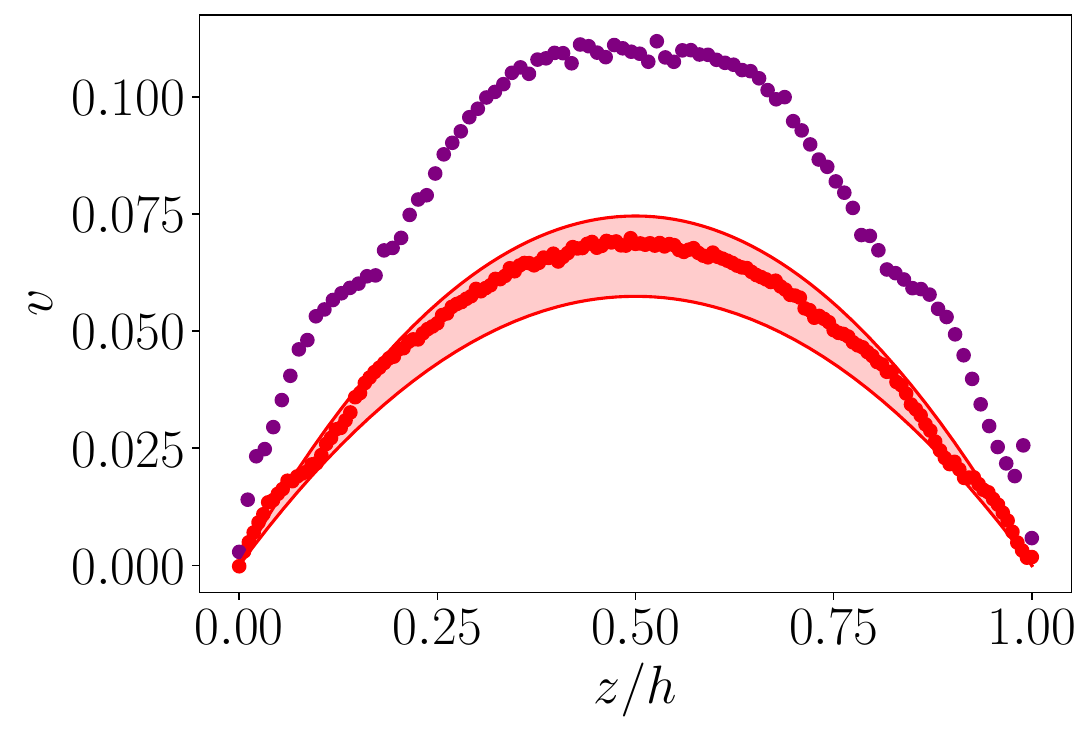}
        \put(-7,70){\makebox(0,0)[lt]{\large\textbf{b)}}}
    \end{overpic}

	\caption{\label{fig:poisprof}
		Poiseuille flow: (a) Shear pressure profiles for $h=7.8$ and $h=3.8$ obtained from the density and acceleration amplitudes and Eq. (\ref{eq:shearPressGen}). (b) Corresponding velocity profiles.
		Shaded area shows the predictions from the NS equation, where uncertainty in the constant viscosity is
		included; no fitting is performed. The viscosity is interpolated from Rowley and Painter \cite{rowley:jtp:1997}. 
	}
\end{figure}

Figure \ref{fig:poisprof} (a) shows the shear stress profiles using the cosine series representation and Eq. (\ref{eq:shearPressGen}). For $h=3.8$ the effect of the additional mode seen in the inset of 
Fig. \ref{fig:pois} is clear, whereas for $h=7.8$ the shear stress follows the linear 
prediction from Newton's model in the bulk. Up till now the discussion has been based on equilibrium density profiles 
and the analytical expression for the amplitudes for the external acceleration. Figure \ref{fig:poisprof} (b)
shows the velocity profiles obtained directly from simulations 
in the presence of an applied acceleration. 

Except very close to the wall the NS equation correctly predicts the flow profile 
for $h=7.8$. On the other hand, for $h=3.8$ the profile features additional modulations
with a wave length of approximately one Lennard-Jones particle diameter; also see 
Travis et al. \cite{travis:pre:1997}.  This is in very good agreement with the spectra 
for the shear pressure.

\begin{figure}
    \centering
    \begin{overpic}[scale=0.5]{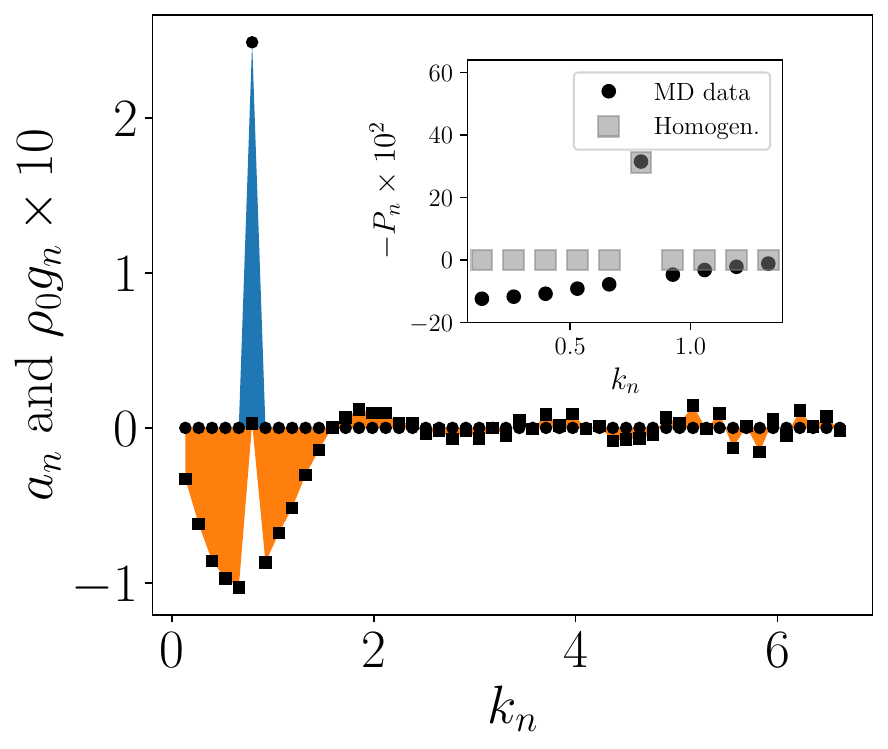}
        \put(-7,80){\makebox(0,0)[lt]{\large\textbf{a)}}}
    \end{overpic}

    \begin{overpic}[scale=0.41]{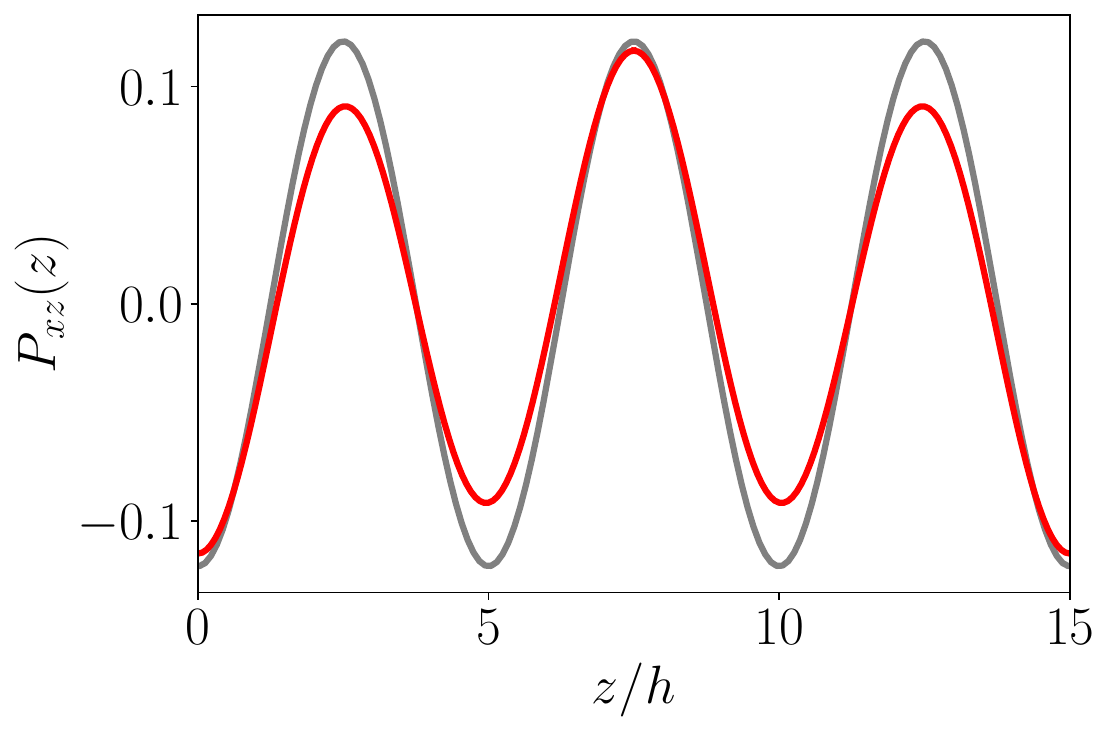}
        \put(-7,70){\makebox(0,0)[lt]{\large\textbf{b)}}}
    \end{overpic}

    \caption{(a) Spectra for the homogeneous part, $\rho_0 g_n$, and inhomogeneous part, $a_n$, for liquid butane with a sinusoidal external force where $n_g=6$. (b) Corresponding shear pressure profile. The gray line shows is the homogeneous case. 
    \label{fig:butane}}
\end{figure}

 
\textit{Sinusoidal transverse acceleration driven flow} 
Next we consider the case where $g_\text{ext}= G_0 \sin(k_g z)$, $k_g$ being the external acceleration 
wave vector $k_g = n_g \pi/h$. While this acceleration is difficult to realize experimentally, it 
enables us to investigate the hydrodynamic mode coupling in a controllable manner using only a single 
mode for $g_\text{ext}$. 

We have $g_n=G_0$, for $n=n_g$ and zero otherwise. For the homogeneous part we then simply have  
$\rho_0 g_n = \rho_0 G_0$. The inhomogeneous part, Eq. (\ref{eq:an}), we write as
\begin{align}
    \label{eq:ansplit}
	a_n = 
		\begin{cases}	
			(\rho_{n_g-n} - \rho_{n_g+n}) g_{n_g} & \text{for } n_g > n \\
			-\rho_{2n_g} & \text{for } n_g = n \\
			( \rho_{n-n_g} - \rho_{n_g+n}) g_{n_g} & \text{for } n > n_g 
		\end{cases}
\end{align}
Let $n_g=6$. In this case, the homogeneous part is zero for wave numbers $n=1, \ldots, 5$,  and from Eq. (\ref{eq:ansplit}) one readily sees that the inhomogeneous part is non-zero. Thus, we have the opposite phenomenon, namely, that the inhomogeneous part dominates the shear pressure for small wave vectors, or on longer length scales. 

Figure \ref{fig:butane} (a) shows the homogeneous and inhomogeneous amplitude spectra for the butane liquid system  driven by a single-mode sinusoidal acceleration with $n_g=6$. The density profile is found in supplementary material. As pointed out above, we see a significant small wave vector inhomogeneous part, whereas the homogeneous part is zero for these small wave vectors. Contrary to the Poiseuille flow, the shear stress here features additional large wave length modulations. In Fig. \ref{fig:butane} (b) we plot the shear pressure clearly featuring these superimposed large wave length modes. Importantly, this phenomenon is not captured by neither generalized hydrodynamics or the LADM. This explains the findings reported in Knudsen et al. \cite{knudsen:pof:2025} and Heitmeier et al. \cite{heitmeier:pof:2025}. 

In conclusion, using eigenfunction decomposition for the density and acceleration we have been able to decompose each  shear pressure applitude into a homogeneous part and an inhomogeneous part given by the density and acceleration Fourier amplitudes. This enables a detailed investigation of the effect of density variations in nanoscale fluid flows. We first revisited the Poiseuille flow and showed the homogeneous part dominates the small wave vector modes as long as the channel height is larger than the characteristic intermolecular distance. This analysis in agreement with previous simulation results. We then showed that by exciting specific acceleration modes, one observes the contrary phenomenon and the inhomogeneous part dominates the flow behavior for long wave lengths. Thus, the effect of density variations on the flow depend not only on the density profile, but also on the acceleration; specifically, how the modes for the two couple.       

\emph{Acknowledgments} This work was supported by the Independent Research Fund Denmark through grant no. 5281-00148B.
\bibliography{lit_all.bib}

\end{document}


\title{Supplemental Material to \textit{Hydrodynamic Mode Coupling: Effects of density variations in nano-scale channel flows.}}

\date{\today}
\author{Linnea Heitmeier}
\email{heitmeier@ruc.dk}
 \affiliation{Institute of Frontier Materials on Earth and in Space, German Aerospace Center, Cologne, Germany}
  \affiliation{Department of Physics, Heinrich-Heine Universität Düsseldorf, Universitätsstraße 1, 40225 Düsseldorf, Germany}
  \affiliation{``Glass and Time'', IMFUFA, Department of Science and Environment, Roskilde University, Postbox 260, DK-4000 Roskilde, Denmark
}
\author{Jesper S. Hansen}
 \email{jschmidt@ruc.dk}
\affiliation{``Glass and Time'', IMFUFA, Department of Science and Environment, Roskilde University, Postbox 260, DK-4000 Roskilde, Denmark
}

\maketitle

\section{Simulation details}
In the paper \textit{Hydrodynamic Mode Coupling: Effects of density variations in nano-scale channel flows} we show two types of simulations. All simulations were carried out with the open-source codes \texttt{gamdpy} \cite{gamdpy} and \texttt{molsim} \cite{molsim}. The latter is used for the simple Lennard-Jones particle simulations, where all particles interact via the standard Lennard-Jones potential 
\begin{align}
     U(r_{ij}) = 4\epsilon \left( r_{ij}^{-12} -  r_{ij}^{-6}\right) \ ,
\end{align}
if $r_{ij} \leq 2.5$, where $r_{ij}$ is the distance between particle $i$ and $j$. If $r_{ij} > 2.5$ the particles do not interact. The wall particles are also tethered bcc lattice sites using a restoring Hookean spring force with spring constant $k=300$. The wall number density is set to $\rho =1$ (in simulation units) and the fluid density is approximately 0.75, estimated from the density profile. The channel dimensions are $10 \times 10 \times h$, where $h=7.8$ or $3.8$. The wall consists of at least four particle layers and acts both as lower and upper wall due to the periodic boundaries. Moreover, the wall particles are connected to a simple relaxation thermostat, see Sadus \cite{sadus:book:1999}, which ensures an average constant temperature in the system of $T=1.25$ (in simulation units), that is, the state point of the confined fluid corresponds to a liquid. 

We also perform simulations of confined butane using a flexible version of the Ryckaert-Bellemans model for alkanes \cite{ryckaert:fdcs:1978, hansen:molsim:2021}. Besides interacting via the Lennard-Jones potential, the model includes covalent bond potential, angle bending, and dihedral interactions between the four united atomic units (uau) defining the molecule, also see \cite{puscasu:condmat:2010} for details. Table \ref{tab:parameters} lists the potential parameters used in the model.
\begin{table}[h!]
    \centering  
\begin{tabular}{l | l }
        \hline \hline \\
         Quantity &  \text{Value (sim. units)}   \\ \\ \hline  \\
         Density ($\rho$) & 1.477 \\
         Temperature ($T$) & 4.16 \\
         Mass uau ($m$) & 1 \\
         Mass wall particle ($m_{\text{Wall}}$) & 4.3\\
         Bond param. ($L_{\text{bond}}$, $k_\text{bond}$) & 0.4, 33e3 \\
         Angle param. ($k_{\text{ang}}$, $k_\text{ang}$) & 1.9, 866  \\
         Dihedral param. ($c_n$)& 15.500, 20.305
         -21.917, -5.115, 43.834, -52.607 \\
        \hline \hline
    \end{tabular}
    \caption{Parameters which were used in the butane simulations. The united atomic unit represents either a -CH$_3$ or a -CH$_2$ group and is given mass 1 in simulation units. The wall particles are chosen to have a mass similar atoms.}
    \label{tab:parameters}
\end{table}

The system is set up in the following way: First, we set $N=8000$ butane molecules on a regular grid, then we compress it, until it has the desired density. This system is then equilibrated in the NVE-ensemble (constant particle number, volume and energy). 
Independently from that, we prepared an amorphous wall by equilibrating a simple Lennard-Jones fluid at $T=2.0$ for $500 \tau$ (details of this model can be found above). We checked that the resulting material is amorphous by studying the corresponding pair-correlation function.  

After that, we combine the two systems by placing the molecules next to the wall, such that the system we simulate is a nanochannel with width $L_z$. Also, from this step on, the wall particles are tethered with a restoring Hookean spring force with a spring constant of $k=500$. 

\begin{figure}[h!]
    \centering
    \includegraphics[width=0.5 \linewidth]{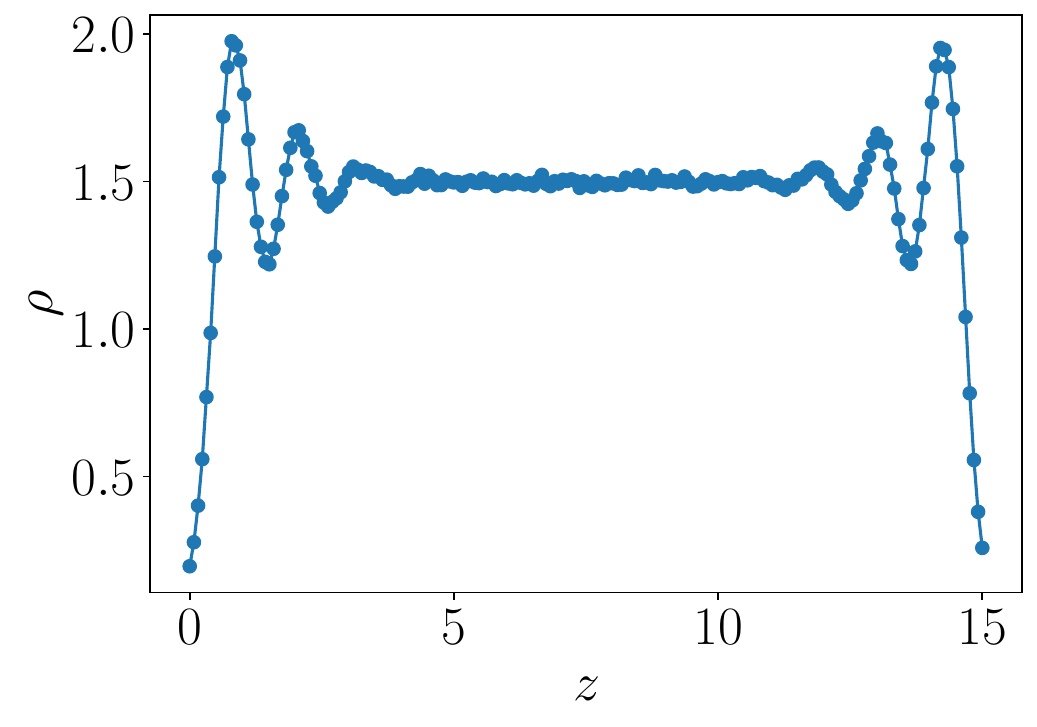}
    \caption{Density profile of butane}
    \label{fig:butane_densityprofile}
\end{figure}

After the system is set up, an external acceleration $g_\text{ext} = A \sin(k_n z)$ is applied to the particles and the density and velocity profiles are sampled using a standard bin method
\[
\rho_\text{bin} =  \frac{1}{V_\text{bin}} \left \langle \sum_{i \in \text{bin}} m_i  \right \rangle \ \text{and} \  
u_\text{bin} = 
\frac{ \left \langle \sum_{i \in \text{bin}} m_i v_{i,x}\right \rangle }
{\langle \sum_{i \in \text{bin}} m_i\rangle}
\]
where $\langle \ldots \rangle$ indicates sample average. The value of $z$ is then chosen as bin mid-point. 
We checked that the values of $A\in(0.05, 0.1)$ were small enough to be in the linear regime. An exemplary density profile can be seen in figure \ref{fig:butane_densityprofile}. 

\bibliography{lit_all.bib}